\documentstyle[11pt,paspconf]{article}

\begin{document}

\title{Mrk~421: {\it EUVE} observations from 1994 to 1997}

\author{I. Cagnoni} 
\affil{International School for Advanced Studies (SISSA-ISAS), Trieste, Italy}

\author{A. Fruscione}
\affil{Harvard-Smithsonian Center for Astrophysics, Cambridge, MA, USA}

\author{I. E. Papadakis}
\affil{University of Crete, Greece}

%\begin{abstract}
%We briefly present 
%spectral and timing analysis of $\sim 1$~Ms of data 
%collected by the {\it Extreme Ultraviolet Explorer} for the BL Lac 
%object Mrk 421 from 1994 to 1997. Further details can be found Cagnoni et al. 1998.
%\end{abstract}

% Keywords should be included, but they are not printed in the hardcopy.


\section{Observations: variability and spectra}

From 1994 to 1997  Mrk~421 was observed by {\it EUVE} (0.02--0.21 keV)
 4  times with the Deep Survey/Spectrograph, 
which allows simultaneous imaging and spectroscopy (e.g. Welsh et al. 1990)
and 2 times with the scanner (imaging) telescopes  
for a total of $\sim 1000$ ks (see Table~1 for details).

\noindent \underline {Variability:} 
Not only {\it EUVE} lightcurves taken from 1994 to 1997 
play a key role in multiwavelenghts studies,
but, since they represent the longest continuous coverages for this source 
at these energies, they allow a detailed variability analysis.
We analyzed all 5 lightcurves in an homogeneous way and we studied the variability
of Mrk~421 through the power spectrum technique.
We fitted the power spectra with a power law model and the best fit slope found  
using all the DS lightcurves binned over
5544 s is $-2.52 \pm 0.40$, while combining the DS and scanner lightcurves
we obtain  $-2.18 \pm 0.30$ (further details can be found in Cagnoni et al. 1998).

\begin{table}
\caption{{\it EUVE} Observations of Mrk~421 from 1994 to 1997}
\vspace{-0.1in}
\begin{center}\scriptsize
\begin{tabular}{l l l l c c c c} 
\hline\hline
\multicolumn{4}{c}{Observations}&
\multicolumn{2}{c}{Lightcurves}&
\multicolumn{2}{c}{Energy Spectra}\\
\hline
Year 	&Instr. &Begin &End   &T$_{\rm exp}$  &Count rate &Slope
 &Normalization\\
&&&&(ks)&($10^{-1}$ c s$^{-1}$) & &($\mu$Jy at 80\AA)\\
\hline
\\
94		&DS/S	 &Apr. 2	&Apr. 12    	&280$^a$ 	&$1.758	\pm 0.009$ 	&$1.0^{+0.6}_{-0.9}$$^b$	&$340 \pm 18$\\ 
95		&Scanner &Feb. 4 	&Feb. 7		&68 		&$3.219 \pm 0.029$$^c$	&--			&--\\ 
95		&DS/S	 &Apr. 25	&May 13		&355		&$2.914 \pm 0.010$	&$1.1 \pm 0.4$		&$355 \pm 12$\\
95(p. 1)	&DS/S	 &Apr. 25 	&Apr. 28 	&86$^d$ 	&$3.743 \pm 0.023$	&$1.4 \pm 0.6$		&$540^{+35}_{-20}$\\ 
95(p. 2)	&DS/S	 &Apr. 28	&May 6		&154$^d$	&$2.950 \pm 0.015$	&$2.2 \pm 0.5$$^e$	&$385^{+16}_{-19}$\\
95(p. 3) 	&DS/S	 &May 7         &May 13         &115 		&$2.240 \pm 0.016$	&$-0.6^{+1.0}_{-1.2}$	&$185^{+21}_{-17}$\\
96		&DS/S	 &Apr. 17	&Apr. 30	&299 		&$3.040 \pm 0.011$	&$2.1 \pm 0.3$		&$495 \pm 15$\\ 
96		&DS/S    &May 10        &May 11         &3.6   		&$4.066 \pm 0.113$	&--			&--\\ 
97		&Scanner &Feb. 7 	&Feb. 11 	&108 		&$1.920 \pm 0.016$$^c$	&--			&--\\ 
       		&        &              &               &...... 	&	&			&\\
		&	 &		&Total		&1114		&	&			&\\
\end{tabular}
\end{center}
\vspace{-0.32in}
\tablenotetext{a,b}{Lightcurve presented in Fruscione et al. 1996; the best fit of the
spectral slope was $1.4 \pm 0.8$}
\tablenotetext{c}{Count rate in the Scanner; the equivalent in the DS/S would be $4.506 \pm 0.041$ in 1995 and $2.690 \pm 0.022$ in 1997}
\tablenotetext{d,e}{Lightcurve presented in Kartje et al. 1997; the best fit of the
spectral slope was $3.5 \pm 0.8$}
\vspace{-0.1in}
\end{table}

\noindent \underline {Energy spectra:} We analyzed the three spectral data sets in an
 homogeneous way using the proper off-axis calibrations
(Marshall 1998, private communication).
We modeled the EUV spectrum of Mrk~421 over the range
 where the ISM  does not absorb all the photons (70-100 \AA; 75-110 \AA
\/ for 1994 observation)
with an absorbed power law whose best fit values are given in Table~1.
In order to check for the presence and depth of the absorption feature claimed by
Kartje et al. 1997 and in an
attempt to follow the development of the flare, we divided the 1995 observation into
three parts (see Table~1): April 25 - 28 (the flare), April 28 - May 6 
(where the absorption feature was claimed) and May 7 - 13.

\section{Discussion}

Unfortunately the large errors on the slope and normalization make it impossible to
follow the spectral development of the flare and even if both
the correction for the proper off-axis effective area and the flattening of 
Mrk~421 spectrum below 1.5 keV of $\sim 0.5$ (Takahashi private communication)
combine to weaken the evidence of the feature around $\sim 70$ \AA,
nothing conclusive can be said.
The total EUVE light curve seems to be 
smoothly varying on the long time-scale while on a shorter 
time-scale there is evidence of an {\it EUVE} flare correlated to 
the 1995 TeV flare.
The best fit value for the slope of the power spectrum
 is consistent with results found for
PKS~2155-304 in the X-rays ($\alpha = -2.5 \pm 0.2$ Tagliaferri et al 1991)
and in optical ($\alpha = -2.4$  Paltani et al 1997);
it is also consistent with the $\alpha = -2$ 
expected from a superposition of exponentially decaying/rising shots.
A comparison with the jet emission inhomogeneous models (e.g. Celotti et al. 1991)
suggests that the shots could be due to perturbations of growing thickness 
propagating outward along the jet.
These perturbations act as an increase of the relativistic electron density 
and magnetic field strength in the emitting region.
Since we do not see any break in the power spectrum, we can place only a lower limit
on the decaying time scales of the shots which are likely responsible for the 
source variability. The limit implies a long 
decaying time scale ($> 1/10^{-5.5}$ Hz$ \simeq 3.66$ days) spread over a narrow 
range.
% (otherwise we would expect detectable features in the power spectrum).

\acknowledgments

This work was supported by AXAF S.C. NASA Contract NAS 8-39073 and by NASA
grant NAG 5-3174 and NAG 5-3191.

\end{document}